\documentclass[epj,twocolumn]{webofc}
\usepackage[varg]{txfonts}   
\usepackage{graphicx}

\wocname{epj}
\woctitle{Seismology of the Sun and the Distant Stars 2016}
\begin{document}
\title{Extended Aperture Photometry of K2 RR Lyrae stars}
%

\author{\firstname{Emese} \lastname{Plachy}\inst{1}\fnsep\thanks{\email{eplachy@konkoly.hu}}, 
        \firstname{P\'eter} \lastname{Klagyivik}\inst{1},
        \firstname{L\'aszl\'o} \lastname{Moln\'ar}\inst{1}, 
        \firstname{\'Ad\'am} \lastname{S\'odor}\inst{1} 
        \and
        \firstname{R\'obert} \lastname{Szab\'o}\inst{1}   
}

\institute{Konkoly Observatory, MTA CSFK, Konkoly Thege Mikl\'os \'ut 15-17, H-1121 Budapest, Hungary}

\abstract{%
We present the method of the Extended Aperture Photometry (EAP) that we applied on K2 RR Lyrae stars. Our aim is to minimize the instrumental variations of attitude control maneuvers by using apertures that cover the positional changes in the field of view thus contain the stars during the whole observation. We present example light curves that we compared to the light curves from the K2 Systematics Correction (K2SC) pipeline applied on the automated Single Aperture Photometry (SAP) and on the Pre-search Data Conditioning Simple Aperture Photometry (PDCSAP) data.}
\maketitle
%

\begin{figure*}
\centering
\includegraphics[width=\hsize,clip]{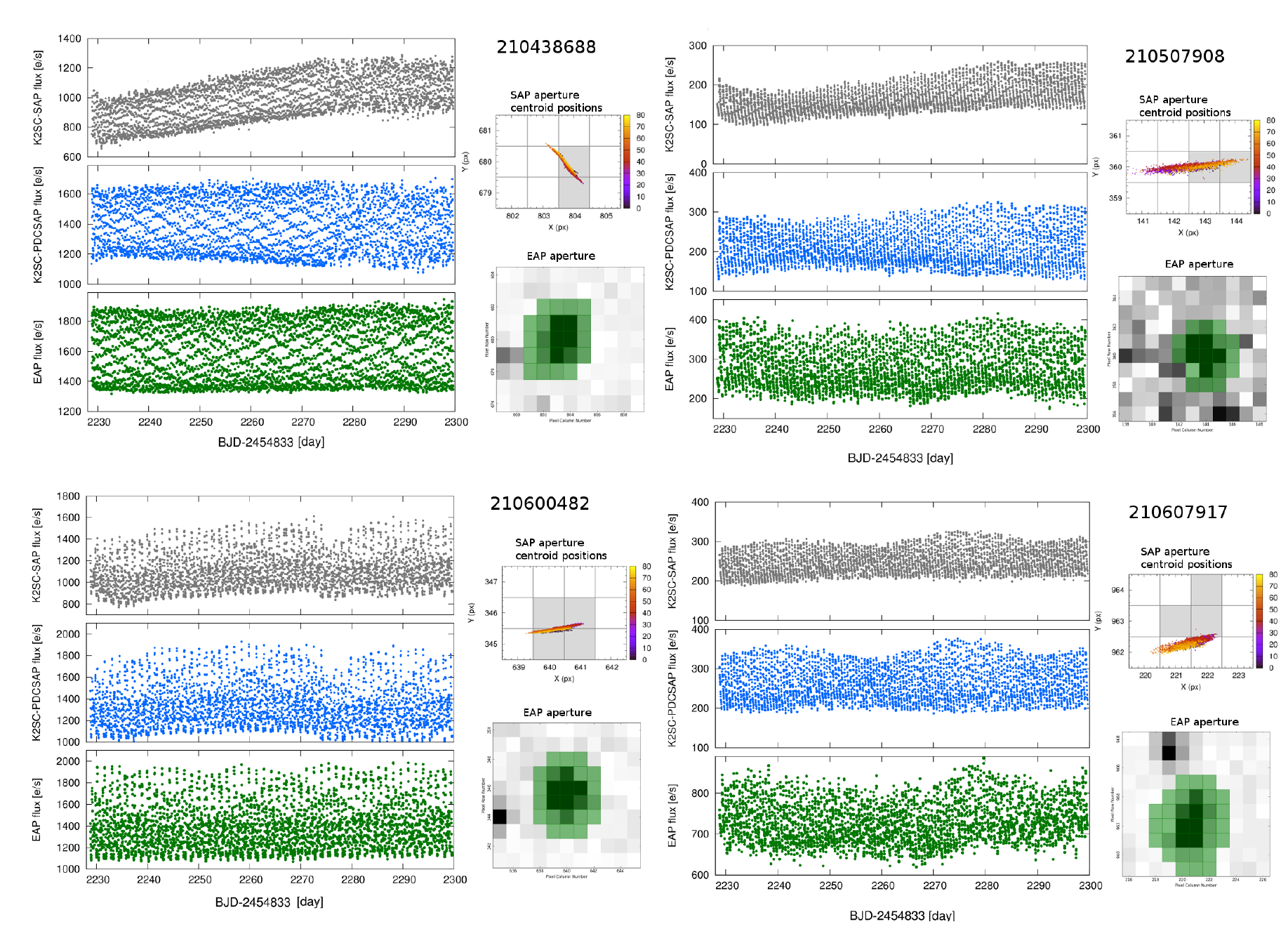}
\caption{Different photometric solutions for EPIC 210438688, EPIC 210600482, EPIC 210507908, EPIC 210607917 are presented. The upper panels (grey) show the K2SC-corrected SAP light curves. Middle panels (blue) show the K2SC and PDC corrected light curves. Lower panels display the EAP photometry. The applied SAP aperture with the PSF centroid positions are on the small panels on the right side of the light curve panels. Colour bars indicate the time evolution of the position in days. The green pixels on the lower small panels show the aperture used in the EAP photometry.}
\label{fig-1}       
\end{figure*}
\section{The need for custom photometric pipelines}
\label{intro}

K2 observations suffer from various instrumental effects. The most serious one is the variation in spacecraft attitude that is corrected on a 6-hour timescale. Various methods have been developed to tackle these issues but neither is really optimal for RR Lyrae stars. Main characteristics of RR Lyrae photometry are the $\sim$6-12 hour periods (in the range of the correction frequency) and sharp features in the light curve (false positives for outliers). Also, we are interested in the full-amplitude variation, and not in a detrended (systematics-corrected) light curve that other pipelines offer. The KepSFF pipeline \cite{van}, for example, is confused by the pulsation, and detrended light curves are of lower quality than the raw fluxes, in several cases. The K2SC pipeline \cite{k2sc} removes the position-dependent systematics from the SAP and PDCSAP light curves. PDCSAP data are already free from systematic trends removed by the PDC (Pre-search Data Conditioning) method \cite{pdc}.

Here we compare our solution for RR Lyrae stars with the outputs of K2SC pipeline. In many cases, EAP photometry retains the full amplitude much better than SAP/PDCSAP-based light curves, but it is sensitive to contamination from blended sources.

\section{A simple way: Extended Aperture Photometry (EAP)}
\label{sec-1}

We created new apertures for every star. The apertures were defined to contain the movements of the stars within the target pixel masks, but to avoid contamination from nearby stars whenever possible. Although this method produces a very slightly increased noise from the background pixels, in many cases it also preserves the pulsation amplitudes throughout the campaign much better than the SAP and PDCSAP fluxes. 

After the photometry, we removed all points with SAP\_QUALITY flags larger than 0, except for C2, where we also kept the large number of points with flags value of $2^{15}$ (16384). 

We then ran an automated Fourier analysis script on the light curves to construct an initial fit for sigma clipping \cite{sodor}. The fit was subtracted from the data, all >3 sigma outliers from the residuals were removed, then the fit was added back.

\section{Examples}
\label{sec-2}

Figure~\ref{fig-1} displays four examples of RR Lyrae stars from Campaign 4.

\subsection{EPIC 210438688 $({Kp} = 17.387 ~\mathrm{ mag})$}

 The aperture used by the Kepler pipeline is clearly undersized. The SAP  data indicates that the PSF periodically moved out of the SAP aperture during the second half of the campaign, leading to significant flux loss. 
The PDCSAP and the K2SC-corrected light curves recovered some of the intrinsic variation, but at the cost of increased scatter. 
In contrast, our EAP photometry collected the flux troughout the campaign, and provided us with an improved light curve for this first-overtone RR Lyrae star.

\subsection{EPIC 210600482 $({Kp} = 17.636 ~\mathrm{ mag})$}

The SAP/PDCSAP fluxes experienced flux loss in the case of this faint RRd star too. The K2SC correction recovered most of the variation, but the last third of the data has larger scatter, and also changes in the average flux level. The EAP light curve nicely follows the double-mode variation along the campaign.

\subsection{EPIC 210507908 $({Kp} = 19.085 ~\mathrm{ mag})$}

This star suffers from contamination from nearby sources. 
Interestingly, both the original PDCSAP and the systematics-corrected one shows an apparent amplitude change that follows the average flux changes of the SAP flux. The EAP flux shows small changes in the average brightness (implying contamination), but no such amplitude variation. We noticed this behaviour in other stars too. This effect could mimic the Blazhko modulation in stars where it is not present.

\subsection{EPIC 210607917 $({Kp} = 18.898 ~\mathrm{ mag})$}

Both the SAP and PDCSAP fluxes suggest the presence of amplitude modulation (Blazhko effect) in this RRab star. The EAP light curve shows much higher flux and strong contamination from a blended source, but no amplitude variation. Here the photocenter tracks the brighter, non-RR Lyrae star.

\section{Conclusions}

Our study revealed a distortion effect of the PDC pipeline on K2 RR Lyrae light curves. When the star moves in and out from the undersized mask a significant amplitude change occurs. After removing the systematics by extracting the Cotrending Basis Vectors, the amplitude variation becomes more prominent and mimics the Blazhko effect. This is very typical when contaminations play a role. On the other hand EAP light curves contain all the fluxes from the blended stars and their variability may be hard to distinguish from the RR Lyrae pulsation. Taken all these together we believe that EAP gives better light curves solutions for RR Lyrae stars than other available methods. However, we note that blended stars require careful analysis.
\label{sec-con}

\section*{Acknowledgements}
\label{sec-ack}
This work has used K2 targets selected and proposed by the RR Lyrae and Cepheid Working Group of the Kepler Asteroseismic Science Consortium (proposal number GO4069). Funding for the Kepler and K2 missions is provided by the NASA Science Mission directorate. This project has been supported by the LP2014-17 Program of the Hungarian Academy of Sciences, and by the NKFIH K-115709, PD-116175 and PD-121203 grants of the Hungarian National Research, Development and Innovation Office. L.M. and \'A.S. were supported by the János Bolyai Research Scholarship of the Hungarian Academy of Sciences. The research leading to these results has received funding from the European Community's Seventh Framework Programme (FP7/2007-2013) under grant agreement no. 312844 (SPACEINN). The authors also acknowledge support from to ESA PECS Contract No. 4000110889/14/NL/NDe. This work made use of PyKE \cite{still}, a software package for the reduction and analysis of Kepler data.

%
%

\end{document}